\documentclass[aps,prl,twocolumn,showpacs,superscriptaddress]{revtex4-1}
\usepackage{graphicx,xcolor} 
\usepackage{dcolumn,multirow} 
\usepackage{bm,amssymb,amsmath,soul} 

\usepackage{scalerel,tikz}
\usetikzlibrary{svg.path}
\definecolor{orcidlogocol}{HTML}{A6CE39}
\tikzset{orcidlogo/.pic={
 \fill[orcidlogocol] svg{M256,128c0,70.7-57.3,128-128,128C57.3,256,0,198.7,0,128C0,57.3,57.3,0,128,0C198.7,0,256,57.3,256,128z};
 \fill[white] svg{M86.3,186.2H70.9V79.1h15.4v48.4V186.2z}
 svg{M108.9,79.1h41.6c39.6,0,57,28.3,57,53.6c0,27.5-21.5,53.6-56.8,53.6h-41.8V79.1z M124.3,172.4h24.5c34.9,0,42.9-26.5,42.9-39.7c0-21.5-13.7-39.7-43.7-39.7h-23.7V172.4z}
 svg{M88.7,56.8c0,5.5-4.5,10.1-10.1,10.1c-5.6,0-10.1-4.6-10.1-10.1c0-5.6,4.5-10.1,10.1-10.1C84.2,46.7,88.7,51.3,88.7,56.8z};
}}
\newcommand\orcidicon[1]{\href{https://orcid.org/#1}{\mbox{\scalerel*{
\begin{tikzpicture}[yscale=-1,transform shape]
\pic{orcidlogo};
\end{tikzpicture}
}{|}}}}

\usepackage[breaklinks,colorlinks,citecolor=blue]{hyperref}
\def\apj{ApJ}
\def\mnras{MNRAS}
\def\nat{Nat}

\def\araa{ARA\&A}
\def\aap{A\&A}

\def\na{NA}

\def\prd{Phys. Rev. D}

\def\jcap{Journal of Cosmology and Astroparticle Physics}

\newcommand{\msun}{{\rm M}_\odot}

\begin{document}

\title{First star-forming structures in fuzzy cosmic filaments}
\author{Philip Mocz~\orcidicon{0000-0001-6631-2566} }
 \thanks{Einstein Fellow}
 \email{pmocz@astro.princeton.edu}
 \affiliation{Department of Astrophysical Sciences, Princeton University, 4 Ivy Lane, Princeton, NJ, 08544, USA}
\author{Anastasia Fialkov~\orcidicon{0000-0002-1369-633X}}
 \affiliation{Institute of Astronomy, University of Cambridge, Madingley Road, Cambridge CB3 0HA, UK}
 \affiliation{Kavli Institute for Cosmology, University of Cambridge, Madingley Road, Cambridge CB3 0HA, UK}
 \affiliation{Department of Physics and Astronomy, University of Sussex, Falmer,  Brighton BN1 9QH, UK}
\author{Mark Vogelsberger~\orcidicon{0000-0001-8593-7692}}
\affiliation{Department of Physics, Kavli Institute for Astrophysics and Space Research, Massachusetts Institute of Technology, Cambridge, MA 02139, USA}
\author{Fernando Becerra~\orcidicon{0000-0002-8282-4024}}
\affiliation{Harvard-Smithsonian Center for Astrophysics, 60 Garden Street, Cambridge, MA 02138, USA}
\author{Mustafa A. Amin~\orcidicon{0000-0002-8742-197X}}
\affiliation{Physics \& Astronomy Department, Rice University, Houston, Texas 77005-1827, USA}
\author{Sownak Bose~\orcidicon{0000-0002-0974-5266}}
\affiliation{Harvard-Smithsonian Center for Astrophysics, 60 Garden Street, Cambridge, MA 02138, USA}
\author{Michael Boylan-Kolchin~\orcidicon{0000-0002-9604-343X}}
\affiliation{Department of Astronomy, The University of Texas at Austin, 2515 Speedway, Stop C1400, Austin, TX 78712-1205, USA}
\author{Pierre-Henri Chavanis~\orcidicon{0000-0002-2386-3904}}
\affiliation{Laboratoire de Physique Th\'eorique, Universit\'e Paul Sabatier, 118 route de Narbonne 31062 Toulouse, France}
\author{Lars Hernquist~\orcidicon{0000-0001-6950-1629}}
\affiliation{Harvard-Smithsonian Center for Astrophysics, 60 Garden Street, Cambridge, MA 02138, USA}
\author{Lachlan Lancaster~\orcidicon{0000-0002-0041-4356}}
\affiliation{Department of Astrophysical Sciences, Princeton University, 4 Ivy Lane, Princeton, NJ, 08544, USA}
\author{Federico Marinacci~\orcidicon{0000-0003-3816-7028}}
\affiliation{Department of Physics \& Astronomy, University of Bologna, via Gobetti 93/2, 40129 Bologna, Italy}
\author{Victor H. Robles~\orcidicon{0000-0002-9497-9963}} 
\affiliation{Department of Physics and Astronomy, University of California, Irvine, CA 92697, USA}
\author{ Jes\'us Zavala~\orcidicon{0000-0003-4442-908X}}
\affiliation{Center for Astrophysics and Cosmology, Science Institute, University of Iceland, Dunhagi 5, 107 Reykjavik, Iceland}

\date{Aug 30, 2019 \textbf{PRL accepted, Editors' Suggestion}} 

\begin{abstract}
In hierarchical models of structure formation, the first galaxies form in low-mass dark matter potential wells, probing the behavior of dark matter on kiloparsec (kpc) scales. Even though these objects are below the detection threshold of current telescopes, future missions will open an observational window into this emergent world. In this Letter we investigate how the first galaxies are assembled in a `fuzzy' dark matter (FDM) cosmology where dark matter is an ultralight $\sim 10^{-22}$~eV boson and the primordial stars are expected to form along dense dark matter filaments. Using a first-of-its-kind cosmological hydrodynamical simulation, we explore the interplay between baryonic physics and unique wavelike features inherent to FDM. In our simulation, the dark matter filaments show coherent interference patterns on the boson de Broglie scale and develop cylindrical soliton-like cores which are  unstable under gravity and collapse into kpc-scale spherical solitons. Features of the dark matter distribution are largely unaffected by the baryonic feedback. On the contrary, the distributions of gas and stars, which do form along the entire filament, exhibit central cores imprinted by dark matter -- a smoking gun signature of FDM.
\end{abstract}

\maketitle

\begin{figure*}[ht!]
\includegraphics[width=0.97\textwidth]{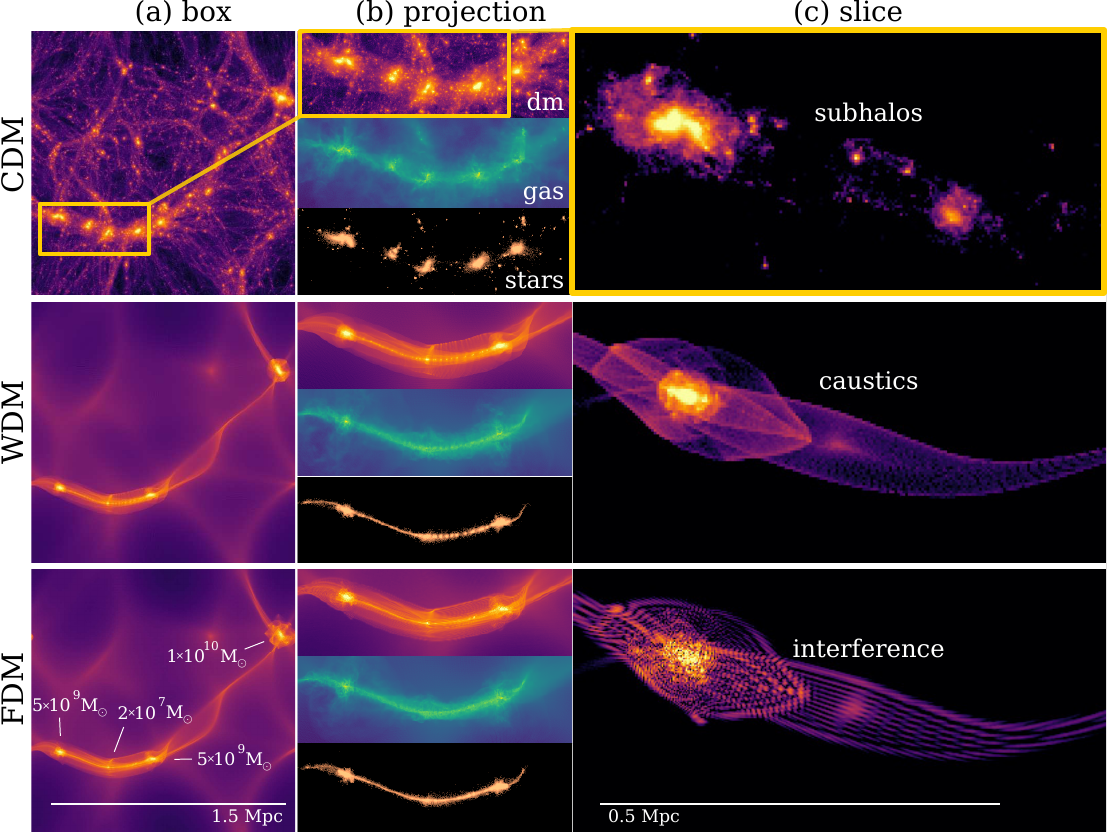}
\caption{Anatomy of a cosmic filament. 
We show, for CDM, WDM, and FDM cosmologies: (a) the projected dark matter distribution in the simulation domain at redshift $z=5.5$; (b) projections of dark matter, gas, and stars in a filament; and (c) slices of the dark matter through a filament.
In CDM the dark matter fragments into subhalos on all scales. 
WDM exhibits rich caustic structures.
FDM has interference patterns at the scales of the de Broglie wavelength, which regularize caustic singularities. 
These differences in small-scale structure will help constrain the elusive nature of dark matter.
}
\label{fig:slice}
\end{figure*}


\begin{figure}[ht!]
\includegraphics[width=0.47\textwidth]{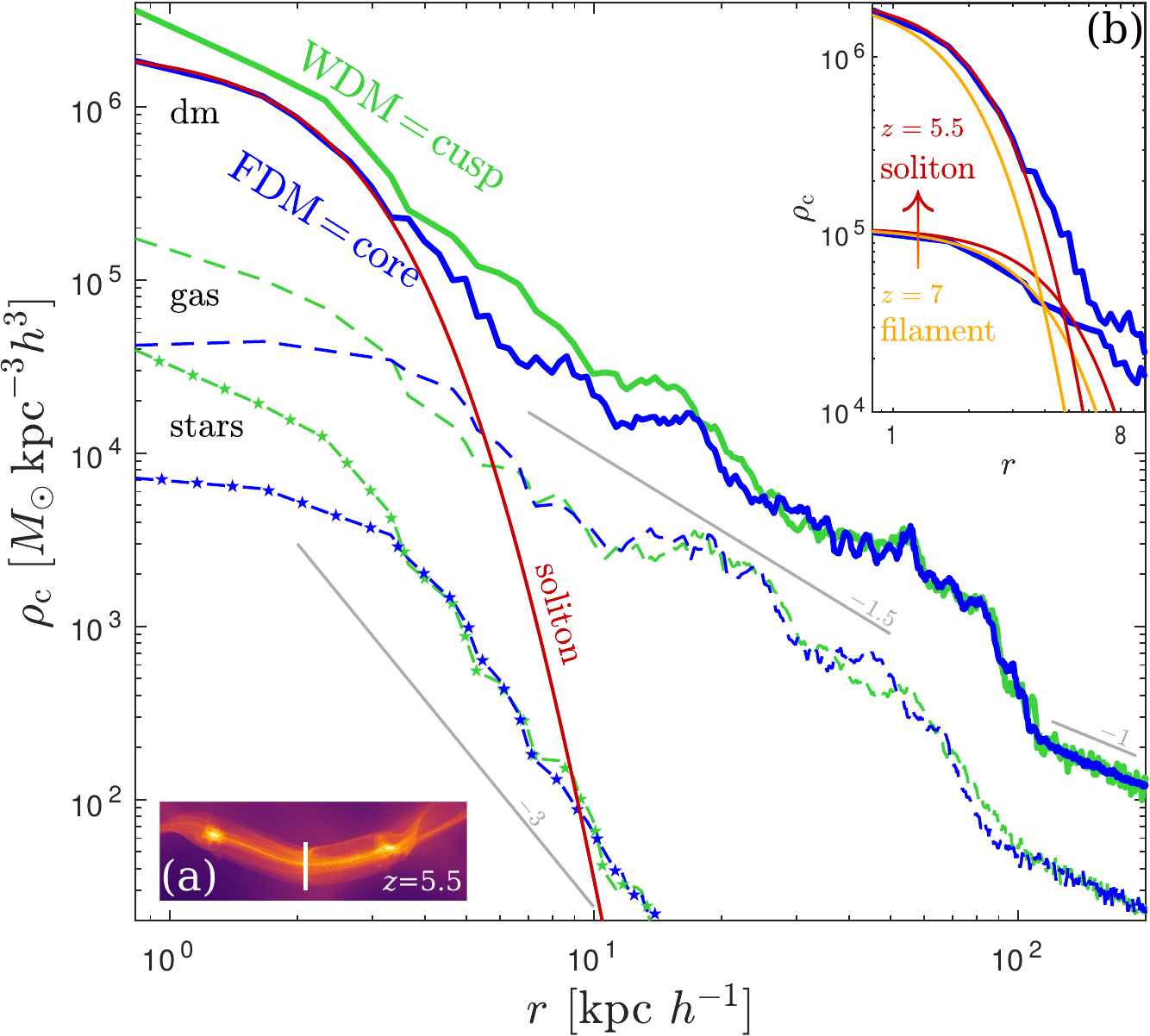}
\caption{The structure of FDM filaments: collapse of a cylindrical filament to a spherical soliton. 
FDM (blue) radial profiles (dark matter\,\,\textemdash\,; gas\,\,--\,--\,; stars\,\,--\,$\star$) are shown through a cross-section of a filament at $z=5.5$ (shown in insert (a)). (b) The dark matter filament has previously ($z=7$) gone unstable from a cylindrical solution and formed a soliton core (the yellow/red lines are cylindrical/spherical profiles of Equation~\ref{eqn:fit}).
Gas traces the dark matter on all scales; while stars form steeper profiles in the filament `spine', but are still cored in the center.
In contrast, WDM (green) exhibits cuspy profiles. 
In CDM (not shown) the filament fragments into multiple subhalos, so the cross-section profile is ill-defined.
Characteristic power law dependencies are shown with gray lines.
}
\label{fig:profile}
\end{figure}


\textit{Introduction}. 
The nearly century-old dark matter problem is one of the most intriguing mysteries in modern physics. We do not know the nature of 84 percent of matter in the Universe, yet it is thought to govern cosmic structure and hold galaxies and clusters together \citep{1933AcHPh...6..110Z}. Observations show that on scales larger than a few megaparsecs (Mpc), the behavior of dark matter is consistent with it being collisionless \citep{PlanckCP:2016,2018MNRAS.475..676S}. However, on scales at and below the size of dwarf galaxies (few kpc)   dark matter is not well constrained \cite{Gilman2019}, allowing for many plausible theories with exotic small-scale physics and particle masses spanning over $30$ orders of magnitude \citep{Hu:2000,2000NewA....5..103G,2000CQGra..17L...9G,2001PhRvD..63f3506M,2009NJPh...11j5027B,2013PhRvD..88d3502V}. The first star-forming regions in the Universe -- more susceptible to dark matter's small-scale behavior than much heavier present-day galaxies -- will be revealed by next generation space telescopes and offer a unique probe of the nature of this elusive component. 

A leading hypothesis for the dark matter `back-bone' of the Universe is cold dark matter (CDM), such as a thermally-produced weakly interacting massive particle (WIMP) of mass $\gg{\rm eV}$. CDM is collisionless and Jeans unstable to forming structure on all astrophysical scales down to a particle physics model-dependent small-scale cutoff (e.g. $\sim$~Earth mass / $10^{-4}$~kpc for a 100 GeV WIMP \citep{2009NJPh...11j5027B})
Apart from a set of ``small-scale controversies'' \citep{2015PNAS..11212249W,2017ARA&A..55..343B}, including the ``cusp versus core'' problem where simulated cuspy halos of galaxies contradict cored observations which may be explained with baryonic effects, CDM has been very successful at describing the observed large-scale structure \citep{PlanckCP:2016,2018MNRAS.475..676S}. 
However, direct and indirect dark matter searches have thus far failed to detect such particles \citep{2018RPPh...81f6201R}.
As a result, there is increased focus on alternative viable scenarios, including warm dark matter (WDM), which is often associated with fermions of particle mass of a few keV (typically treated as collisionless), Peccei-Quinn axions \citep{1977PhRvL..38.1440P} which are bosons of mass $\sim 10^{-5}$--$10^{-3}~{\rm eV}$, and the ultralight FDM of mass $m\sim 10^{-22}$~eV, which is described by a classical scalar field and exhibits wave phenomena on scales of the de Broglie wavelength $\lambda_{\rm dB}\sim {\rm few}\times{\rm kpc}$ \citep{Hu:2000,2000NewA....5..103G,2000CQGra..17L...9G,2001PhRvD..63f3506M,2010PhRvD..81l3530A,Hui:2016}. 
WDM and FDM both yield smoother structures than CDM on scales below few kpc, due to either thermal motion (WDM) \citep{2013PhRvD..88d3502V} or quantum pressure (FDM) \citep{Hu:2000}.
The existence of dwarf galaxies in dark matter halos with masses of $\sim 10^9$ times the mass of the sun ($\msun$) in the local Universe, as well as measurements of the `lumpiness' of the dark matter distribution, constrain WDM and FDM theories, favoring particle masses above $m_{\rm WDM}\sim 3$~keV and $m\sim 10^{-22}$~eV respectively \citep{2017PhRvD..96b3522I,2017PhRvL.119c1302I,2017MNRAS.472.1346G}. The subhalo mass function may imply even higher masses \citep{2019PhRvL.123e1103M}. However, for FDM these constraints can only be used as guidelines, being based on simulations that ignore the impact of wave effects on baryons. 

The first objects in the Universe offer a unique way to tighten the observational constraints.
Compared to the local Universe, in which galaxies in $10^{11}~\msun$ dark matter halos are typical, an early CDM universe (at redshift $z\sim 30$, i.e., $10^8$~years after the Big Bang) is populated by much smaller nearly-spherical halos of $\sim 10^{5}-10^7~\msun$ in which proto-galaxies are born \citep{1977MNRAS.179..541R}. In contrast, WDM first star-forming structures form later and are filamentary due to the initial suppression of the dark matter power spectrum by particle free-streaming \citep{Yoshida:2003,2007Sci...317.1527G}. Compared to WDM, wavelike FDM additionally features interference patterns and soliton cores, as is demonstrated by dark matter-only cosmological simulations \citep{2014NatPh..10..496S}. Until now, impact of FDM on star and galaxy formation has been studied with hydrodynamical
simulations that ignore the wavelike aspects of the dark matter superfluid \citep{Hirano:2017}. The first consistent cosmological simulations of ultralight bosons coupled to the state-of-the-art hydrodynamical modeling are presented here and will allow realistic tests of FDM with existing and upcoming data. 


\textit{Simulating a `fuzzy' universe}. FDM, a scalar boson in the non-relativistic limit, is described by a complex field $\psi = A\exp[-i\phi]$, with amplitude $A$ tied to the dark matter density $\rho\equiv\lvert A\rvert^2$; and phase $\phi$ encoding the velocity $\mathbf{v}\equiv (\hbar/m)\nabla \phi$, where $\hbar$ is the reduced Planck constant.

The Schr\"odinger-Poisson (SP) equations in an expanding universe govern the evolution of FDM \cite{Hu:2000}. In physical coordinates:
\begin{equation}
i\hbar \left(\frac{\partial }{\partial t} +\frac{3}{2}H \right) \psi = -\frac{\hbar^2}{2m}\nabla^2\psi + m V\psi
\label{eqn:schrodinger}
\end{equation}
\begin{equation}
\nabla^2 V = 4\pi G(\rho_{\rm tot}-\overline{\rho}_{\rm tot})
\label{eqn:poisson}
\end{equation}
where $H$ is the Hubble constant, $V$ is the total gravitational potential due to dark matter and baryons, $\rho_{\rm tot}$ is the total density field,
$\overline{\rho}_{\rm tot}$ is the average density of the Universe, and $G$ is the gravitational constant.
The equations approximate the Vlasov-Poisson equations for CDM in the limit of large boson mass or for high halo masses \citep{2018PhRvD..97h3519M}, which makes the study of low mass first structures of particular interest because it is where wavelike effects are expected to be the strongest.

We employ the magneto-hydrodynamics code {\sc arepo} \citep{2010MNRAS.401..791S}, which has been previously used to carry out detailed simulations of galaxy formation with CDM (e.g., the Illustris-TNG project \citep{2018MNRAS.475..676S}). Here, we replace CDM by a FDM formalism via a spectral technique \citep{2017MNRAS.471.4559M}, which evolves the wavefunction in a unitary manner by taking alternating steps to shift the phases of $\hat{\psi}$ (the Fourier transform of $\psi$) to account for the kinetic operator in the Schr\"odinger equation and the phases of $\psi$ itself to account for the gravitational potential. To verify the method we have carried out extensive tests of the convergence of the dark matter power spectrum as a function of resolution and box size. The method requires fixed spatial resolution across the simulated box and fixed (rather than hierarchical) time steps, and, therefore, the wave effects can only be fully explored in small cosmological volumes (up to few Mpc on a side). It is complementary to particle-based methods \cite{2019MNRAS.tmp.1911H,2018MNRAS.478.3935N} which can treat larger box sizes at the cost of not fully-resolving wave interference patterns. The relevant baryonic physics implemented in {\sc arepo} includes sub-grid models for primordial and metal-line cooling, chemical enrichment, star formation, supernova feedback via kinetic winds, and instantaneous uniform reionization at $z\sim 6$ \citep{2013MNRAS.436.3031V}. 
Simulations of FDM were carried out on the TACC \textit{Stampede} supercomputer using $3$~million CPU core hours. The FDM simulations require about $20$ times more computation time than comparison CDM/WDM simulations due to the resolution requirements.

We simulate a volume of size $L_{\rm box}=1.7h^{-1}$~Mpc and assume a boson mass of $m=2.5\times 10^{-22}~{\rm eV}$, which introduces a cutoff in the initial power spectrum at $L_{\rm cutoff}\simeq 1.4h^{-1}$~Mpc due to the uncertainty principle. We evolve the simulation from redshift $z=127$ (Universe age $10^7$ years), where initial conditions are generated using the publicly available Boltzmann code {\sc axionCAMB} \citep{axionCAMB}, to redshift $z=5.5$ (Universe age $10^9$ years), with the final redshift limited by resolution requirements that guarantee fully-converged results. The dark matter spectral resolution is $1024^3$, and the baryon resolution is $512^3$ particles (equivalent to the mass resolution $\sim10^3~\msun$). Cosmological parameters, as measured by the {\it Planck} satellite \citep{PlanckCP:2016}, are assumed, with the exception of $\sigma_8$, which is boosted from $0.8$ to $1.4$ to enhance initial fluctuations and  compensate for the small cosmological volume probed by the simulation (e.g., \citep{Naoz2012}). 

We compare the FDM simulations to those of CDM and WDM which were run with the resolution of $512^3$ dark matter particles
using the same hydrodynamical setup as FDM and the same initial perturbations modulo initial power spectrum shape: CDM has no cutoff, while FDM and WDM assume the same exponentially-suppressed initial power spectrum. Particle masses in FDM and WDM cosmologies can be related by matching the cutoff scale \citep{Hui:2016,Hirano:2017}: the WDM particle mass that corresponds to our choice of $m$ is $m_{\rm WDM} \sim 1.4$ keV. Our WDM is not an exact classical WDM simulation because we ignore initial velocity dispersion of WDM particles. Instead, the WDM case is designed to approximate `FDM minus wave effects' and treats the dark matter as collisionless, an approximation widely used in cosmology (e.g., \citep{Hirano:2017}). 


\textit{First structures in FDM/WDM/CDM}. 
We illustrate the main conceptual differences between the anatomy of the first star-forming structures with CDM, WDM, and FDM in Fig.~\ref{fig:slice} by showing the dark matter, star, and gas distributions across a filament. Fig. \ref{fig:profile}  shows radial profiles for a cross-section perpendicular to the filament.

On large cosmological scales the projected dark matter density fields look similarly smooth in WDM and FDM: the initial suppression in power at $L_{\rm cutoff}$ prevents the formation of halos with masses below $M_{1/2}\simeq 5\times 10^{10}~\msun \left(m/10^{-22}~{\rm eV}\right)^{-4/3}$ \citep{Hui:2016}, and the cosmic web is dominated by dense filaments, which can fragment due to a linear instability to form halos \citep{1997ApJ...479...46V}. In contrast, CDM filaments hierarchically fragment into nearly-spherical subhalos that are resolved down to the simulation mass resolution. 

FDM and WDM strongly differ in their small-scale structure. In WDM, filaments show sharp caustic features in their dark matter distribution (Fig.~\ref{fig:slice}), and the first structures are cuspy (Fig. \ref{fig:profile}). WDM is also known to be susceptible to discreteness noise \citep{2007MNRAS.380...93W} -- i.e., numerical fragmentation of filaments at late times -- due to the lack of a regularizing force, which is seen to an extent in our simulations. 
 In contrast, in FDM caustics are regularized by the uncertainty principle, and structure shows interference patterns from wave superposition. The quantum pressure also prevents the artificial numerical fragmentation seen in WDM. In filaments, the interference remains coherent due to a limited number of wave velocities from the initial collapse, and interference minima/maxima are aligned on scales of ${\rm few}\times 100~{\rm kpc}$. Inside halos, the structure is more complex: waves mimic the multiple shell-crossing in classical collisionless dynamics. Fluctuating kpc-scale wave interference patterns arise, and, gravitationally coupled to baryons, may provide dynamical heating and friction and thicken galactic disks \citep{2019ApJ...871...28B,2019MNRAS.485.2861C}. The size of the interference patterns in filaments and halos is a locally varying quantity, which we find can be estimated from our WDM simulations as the de-Broglie wavelength $\lambda_{\rm dB} = h/(m\sigma)$ of the local velocity dispersion $\sigma$ of the dark matter particles to within a factor of $2$, in line with theoretical predictions \cite{2018PhRvD..97h3519M}. 


On scales of order $\lambda_{\rm dB}$, structures in dense regions can also be highly nonlinear, showing large differences between FDM and WDM. The quantum pressure in FDM can become strong enough to counteract the self-gravity of the dark matter superfluid.
This results in cosmic structures that are unique to FDM such as a spherical soliton core with a radius of few kpc at halo centers \citep{2014PhRvL.113z1302S} versus \textit{much denser} cusps in CDM \citep{1996ApJ...462..563N} and WDM. WDM, though tracing the FDM filament quite well at early times and at large radii, instead collapses into a \textit{denser} cuspy halo due to the absence of quantum pressure support. 
If primordial thermal velocities of WDM (not modeled here) were included, 
the cusp would eventually form a $\sim 10$~pc core by $z\sim 2$ (for our halo masses and effective $m_{\rm WDM}$ particle mass) \cite{2011JCAP...03..024V,2017JCAP...11..017L,2012MNRAS.424.1105M}, which is significantly smaller than the cored structures of FDM formed on kpc scales.

The smallest soliton mass that can form in a cosmological setting as a result of non-linear evolution of the density field is predicted to be $M_{\rm min} \simeq 1.4\times 10^7~\msun \left(m/10^{-22}~{\rm eV}\right)^{-3/2}$ for a boson mass $m$  \citep{Hui:2016}. $M_{\rm min}$ is $3$ orders of magnitude below the minimum mass allowed by the initial power cutoff $M_{1/2}$, which agrees well with halos found in our simulations.  

We find that, in addition to spherical halos, the centers of cylindrical filaments may be supported by quantum pressure (a 2D, unstable version of the 3D soliton). In fact, first non-linear FDM structures seen in our simulations are cylindrical solitons which are unstable and evolve into spherical solitons.
The spherical soliton in halo centers and the cylindrical solution in filaments are well-approximated by: 
\begin{equation}
\rho(r)\simeq
\rho_0
\left[1+\begin{Bmatrix}0.091 & {\rm spherical} \\ 0.127 & {\rm cylindrical} \end{Bmatrix}\times \left(\frac{r}{r_{\rm c}}\right)^2\right]^{-8}
\label{eqn:fit}
\end{equation}
where $r$ is a cylindrical coordinate for filaments and spherical for halos, $r_{\rm c}$ is the core radius and $\rho_0$ is the central density:
\begin{equation}
\rho_0\simeq 1.9\times 10^{9} 
\left(\frac{10^{-22}~{\rm eV}}{m}\right)^2
\left(\frac{{\rm kpc}}{r_{\rm c}}\right)^4
\frac{\msun}{{\rm kpc}^3}.
\end{equation}
The cylindrical filament solution (which we obtained as a fit to the numerical solution for the ground state of the SP equations in cylindrical symmetry) is a squeezed version of the spherical soliton.
Fig.~\ref{fig:profile} (b) shows the radial density profile of a slice through a FDM filament at two cosmic times $z=7$ and $z=5.5$. Initially, the central filament `spine' is well-modeled by the cylindrical filament solution. This first structure is highly triaxial with minor-to-major axis ratio $\sim0.1$. The filament goes unstable and forms a soliton core of mass $M\simeq 2\times10^7~\msun$ by $z=5.5$, near the predicted minimum nonlinear mass limit $M_{\rm min}$.

\textit{Gas and Stars}.
In standard CDM scenarios baryons follow dark matter on scales larger than the filtering scale (the characteristic distance on which pressure acts, e.g., \citep{Gnedin:1998}), while on smaller scale  gas is diffused by pressure. In our FDM and WDM simulations gas pressure does not play a role in the initial collapse of baryonic structure  because the filtering scale  is below the cutoff scale of the initial power spectrum. As a result, the dense FDM/WDM filaments are able to collect gas along the entire structure, in contrast with the fragmentation seen in CDM (Fig. \ref{fig:slice}). In principle, baryons could alter the central dark matter structure in galaxies through gravitational potential fluctuations \cite{2014Natur.506..171P}, but this effect depends on how extended the star formation history is, and is not seen in our simulations. We find that inside filaments the  gas profile traces that of dark matter (up to a lower normalization, Fig. \ref{fig:profile}), and, importantly,  in FDM the dark matter cored-soliton profiles are \textit{imprinted} in the distributions of gas and stars creating potentially detectable smoking-gun signatures of FDM.  

First stars form in deep-enough potential wells which can compensate against pressure and cool gas efficiently.  
Even though supernova feedback and photo-heating of the UV background during reionization may influence the distribution of gas and stars,
the FDM/WDM filaments are dense enough to be \textit{entirely} lit up  by the first generation of stars  even in the presence of sub-grid models for baryonic feedback.  These early star forming filaments  -- structures potentially detectable by the James Webb Space Telescope (JWST) -- are quite different from galaxies we see at later epochs. However, at lower redshifts gravity will fragment these early structures and stars will accrete into more spherical objects, forming familiar-looking galaxies. 

The first FDM galaxies are expected to be intrinsically dimmer than the ones in both WDM and CDM due to the  cored FDM structures and shallower potential wells.  FDM is less efficient in collecting gas and forming stars both in the filaments and halo centers. Approximately $10\%$ less stellar mass is formed by $z=5.5$ compared to  WDM, and $\sim 40\%$ less than CDM (note, however, that in these simulations we  ignore the effect of streaming velocity between dark matter and gas \citep{Tseliakhovich:2010} which might  affect the reference CDM case delaying star formation in small halos of $10^5-10^7$ M$_\odot$ by a few million years \citep{Fialkov:2012}). The suppression in the central stellar density may lead to potentially detectable effects. For instance, inefficient re-population of the orbits along which stars are tidally disrupted by the central black hole will result in fewer tidal disruption events (TDEs). High-redshift TDEs might be observable with next generation transient telescopes if they generate relativistic jets \citep{Fialkov2017, Roming2018}.




\textit{Concluding remarks}.
The first structures that form in the Universe may give away the physical nature of dark matter.
We have carried out first-of-their-kind cosmological simulations of a high-redshift ($z\geq 5.5$) universe with FDM gravitationally coupled to baryons, including sub-grid models for star formation, feedback, and reionization. 
With these simulations we have highlighted systematic differences between FDM, WDM and CDM and explored the interplay between the quantum wave effects in FDM and baryonic physics in the context of first galaxy formation.

First galaxies, targets of JWST, would appear filamentary in both WDM and FDM. The dark matter structure at early times is largely unaffected by the effects of baryonic feedback and reionization, while the distribution of baryons is driven by dark matter even on small scales (kpc). A unique signature of FDM is flattened central profiles in the distribution of gas and stars in halos and filaments, which leads to reduced cosmic star formation. 

Our simulations confirm that on large scales, and even with sub-grid baryonic feedback, the WDM model approximates FDM fairly well, as has been assumed previously \citep{2017PhRvL.119c1302I,Hirano:2017}, suggesting that the Lyman-$\alpha$ forest is a reasonably good tracer of the small-scale dark matter power spectrum even in the nonlinear regime of structure formation.
The discrepancy between FDM and WDM is expected to increase at lower redshifts where quantum pressure wave dispersion acts for longer time.
We have assumed a boson mass $m=2.5\times 10^{-22}~{\rm eV}$, which is in moderate tension with Lyman-$\alpha$ observations \citep{2017PhRvL.119c1302I,2019MNRAS.482.3227N} and the Milky Way subhalo mass function \citep{2019PhRvL.123e1103M}, and future work will test larger boson masses up to $m\sim 10^{-18}~{\rm eV}$ requiring much higher numerical resolution. Such particles would impact formation of the first star-forming objects in halos down to $10^5~\msun$, affecting observable properties of high-redshift galaxies while remaining consistent with the local Universe. 


\begin{acknowledgments}
\textit{Acknowledgements}. We thank Jerry Ostriker, Mariangela Lisanti, David Spergel, Scott Tremaine, James Bullock, Frenk van den Bosch, Paul Shapiro, Jim Stone, Vasily Belokurov, and Aaron Szasz for discussions related to an earlier version of this manuscript. We would also like to thank the anonymous referees who have helped improve the content and presentation of the Letter. The authors acknowledge the Texas Advanced Computing Center at The University of Texas at Austin for providing HPC resources that have contributed to the research results reported within this Letter: \url{http://www.tacc.utexas.edu}; XSEDE Allocation TG-AST170020. Some computations were run on the Odyssey cluster supported by the FAS Division of Science, Research Computing Group at Harvard University. 
Support (P.M.) for this work was provided by NASA through Einstein Postdoctoral Fellowship grant number PF7-180164 awarded by the \textit{Chandra} X-ray Center, which is operated by the Smithsonian Astrophysical Observatory for NASA under contract NAS8-03060. 
A.F. is supported by the Royal Society (URF). 
M.A. is supported by a DOE grant DE-SC0018216, and thanks the Yukawa Institute for Theoretical Physics at Kyoto University, where this work was completed during the YITP-T-19-02 on ``Resonant instabilities in cosmology''.
S.B. acknowledges support from Harvard University through the ITC Fellowship. 
MBK acknowledges support from NSF grant AST-1517226 and CAREER grant AST-1752913 and from NASA grants NNX17AG29G and HST-AR-14282, HST-AR-14554, HST-AR-15006, and HST-GO-14191 from the Space Telescope Science Institute, which is operated by AURA, Inc., under NASA contract NAS5-26555.  
F.M. is supported by the program ``Rita Levi Montalcini'' of the Italian MIUR.
V.H.R. was supported by a Gary A. McCue postdoctoral fellowship.
J.Z. acknowledges support by a Grant of Excellence from the Icelandic Research Fund (grant number 173929-051). 
\end{acknowledgments}

\bibliography{mybib}{}

\end{document}